\documentclass[pdflatex,sn-mathphys-num,iicol]{sn-jnl}

\usepackage{graphicx}
\usepackage{physics}
\usepackage{multirow}
\usepackage{amsmath,amssymb,amsfonts}
\usepackage{amsthm}
\usepackage{mathrsfs}
\usepackage[title]{appendix}
\usepackage{xcolor}
\usepackage{textcomp}
\usepackage{manyfoot}
\usepackage{booktabs}
\usepackage{algorithm}

\usepackage{amsmath}
\usepackage{amssymb}

\usepackage{braket}
\usepackage{varwidth}
\usepackage{algorithmicx}
\usepackage{algpseudocode}
\usepackage{listings}
\usepackage{lineno}
\usepackage[none]{hyphenat}
\usepackage{bbm}
\usepackage{stfloats}

\usepackage{multirow}
\usepackage{lscape}
\usepackage{supertabular,booktabs}

\begin{document}

\title[Article Title]{Simulating the non-Hermitian dynamics of financial option pricing with quantum computers}

\author[1,*]{Swagat Kumar}
\author[1,*]{Colin Michael Wilmott}
\affil[1]{Department of   Mathematics, Nottingham Trent University,  Nottingham, NG11 8NS, UK}
 \affil[*]{swagat.kumar02@ntu.ac.uk, colin.wilmott@ntu.ac.uk}

\abstract{
The Schr\"odinger equation describes how quantum states evolve according to the Hamiltonian of the system. For physical systems, we have it that the Hamiltonian must be a Hermitian operator to ensure unitary dynamics. For anti-Hermitian Hamiltonians, the Schr\"odinger equation instead models the   evolution of quantum states in imaginary time. This process of imaginary time evolution has been used successfully to calculate the ground state of a quantum system. Although imaginary time evolution  is non-unitary, the normalised dynamics of this evolution can be simulated on a quantum computer using the quantum imaginary time evolution (QITE) algorithm. In this paper, we broaden the scope of QITE by removing its restriction to anti-Hermitian Hamiltonians, which allows us to solve any partial differential equation (PDE) that is equivalent to the Schr\"odinger equation with an arbitrary, non-Hermitian Hamiltonian. An example of such a PDE is  the famous Black-Scholes equation that models the price of financial derivatives. We will demonstrate how our generalised QITE methodology offers a feasible approach for real-world applications by using it to price various European option contracts modelled according to the Black-Scholes equation.
}


\maketitle
 A financial derivative is an options contract  whose value derives from an underlying financial asset \cite{Hull17}. An options contract defines an  agreement between two parties that entails a right to trade an  asset at some specified future date for a fixed price. This right to trade agreement  thus creates  inherent value, which   may in turn be traded in the same manner as the underlying financial  asset. Consequently, a financial derivative may be viewed as  an instrument, which can be used to either exploit arbitrage opportunities or mitigate risk exposure in the market. For this reason, a  fundamental task in  quantitative finance is how exactly do we  determine the fair price  of a financial derivative. Determining the fair price of an option is a highly non-trivial task, which is due in part to  the   stochastic nature of the parameters that define a   derivative.  
 
 The famous Black–Scholes model~\cite{BS73, Hull17} is an effective method  for determining the fair price of a derivative, and  has become the standard for pricing European style financial options. Given the payoff price  for an option at the maturity time, we can determine the present price of the option by solving the linear differential equation 
 \begin{equation}
     \frac{\partial u}{\partial t} = -\frac{1}{2}(\sigma x)^2\pdv[2]{u}{x} -rx \frac{\partial u}{\partial x} +ru, \label{eqn:black-scholes.1-eqn}
 \end{equation}
 for $(x,t) \in [x_0,x_N] \times [0,T]$,  where  the  condition $u(x,T) = p(x)$ denotes the payoff of the option. The price of the  option is denoted by $u(x,t)$, while $x$
denotes the value of the underlying asset, $t$ represents time, and $T$ is the maturity time. For simplicity, it is assumed that the volatility of the asset, $\sigma$, and the risk-free interest rate, $r$, are constant with respect to time. For convenience, adopting $\tau = T-t$  transforms the Black-Scholes equation Eq.~(\ref{eqn:black-scholes.1-eqn}) to the initial value problem   
\begin{equation}
     \frac{\partial u}{\partial \tau} = \frac{1}{2}(\sigma x)^2\pdv[2]{u}{x} +rx \frac{\partial u}{\partial x} -ru, \label{eqn:black-scholes-eqn}
 \end{equation}
 for $(x,\tau) \in [x_0,x_N] \times [0,T]$ with the initial  condition $u(x,\tau=0) = p(x)$. To numerically solve the Black-Scholes equation, we must discretise the  domain $[x_0,x_N]$ to a finite domain and assign appropriate boundary conditions. 

The Schr\"odinger equation  models the evolution of the wave function of a quantum mechanical system, and takes the form
\begin{equation}
i \frac{\partial\psi(\vec{x},t)}{\partial t} = \hat{H}\psi(\vec{x},t),    
\label{eqn:schrodinger}
\end{equation}
where the Hamiltonian, $\hat{H}$, is a linear differential operator in $\vec{x}$ acting on the wave function $\psi$. Solutions to the Schr\"odinger equation are expressed in terms of the time evolution operator,
\begin{equation}
    \psi(\vec{x},t) = e^{-i\hat{H}t} \psi(\vec{x},0).
\end{equation}

The Black-Scholes equation, Eq.~(\ref{eqn:black-scholes-eqn}), can also be expressed in the form of the Schr\"odinger equation, where its Hamiltonian is given by 
\begin{equation}
  \hat{H}_{BS} = i\left[\frac{1}{2}(\sigma x)^2\pdv[2]{}{x} + rx \frac{\partial }{\partial x} - r\right].
\label{eqn:Hbs}
\end{equation}
Note that while  the Hamiltonian of the  Schr\"odinger equation is a Hermitian operator, which gives rise to unitary time evolution,  the Black-Scholes Hamiltonian, Eq.~(\ref{eqn:Hbs}), is non-Hermitian, and induces non-unitary time evolution. 
However, since quantum computers evolve under unitary time evolution, it is the case that simulating non-Hermitian dynamics is not directly feasible on a quantum computer. It is for this reason that quantum computing approaches for solving the Black-Scholes equation have thus far been based on variational algorithms \cite{fontanela2021quantum, radha2021quantum, Alghassi2022variationalquantum} or require post-selection techniques \cite{gonzalezconde2022simulating}.

Another example of non-Hermitian dynamics can be seen in the imaginary time evolution of a quantum system. 
Following a Wick rotation, which replaces time with an imaginary number $\beta=it$, the Schr\"odinger equation drives wave functions to become parallel to the ground state of the system. 
The Wick-rotated form of the Schr\"odinger equation also takes the form of Eq.~(\ref{eqn:schrodinger}), but with an anti-Hermitian Hamiltonian. Although the imaginary time Schr\"odinger equation induces non-unitary dynamics, 
the normalised evolution can be simulated with quantum algorithms, including quantum imaginary time evolution (QITE)~\cite{motta2020determining} and variational QITE~\cite{McArdle_2019}.

Variational QITE (varQITE) is a hybrid quantum-classical algorithm that  is well suited for noisy intermediate-scale quantum (NISQ) devices. 
As a variational quantum algorithm, varQITE considers a system of differential equations linking to the gradients of ansatz parameters in imaginary time, and coefficients that  depend on measurements of the ansatz. Variational QITE employs a fixed ansatz, where the time complexity is linear in the number of Hamiltonian terms. However, the choice of ansatz is crucial, as it is possible that the states produced by the true imaginary time evolution may not be generated by the particular parameterised ansatz circuit.

On the other hand, the simulated QITE approach is  an alternative technique  for simulating imaginary time evolution~\cite{motta2020determining}. The technique works by  approximating the  normalised time evolution operator with Trotter products via unitaries. Simulated QITE with sufficiently large unitary domains is not plagued by barren plateaus, as is the case with its  variational counterpart. Simulated QITE on a  $k$-local Hamiltonian requires a number of measurements that is  exponential in $k$,  with the depth of the associated quantum circuits  scaling accordingly. Interestingly, however, recent work has focused on optimising   the    circuit depth and the number of measurements required in simulated QITE. For instance, Fast  QITE  provides for an exponential reduction in the circuit depth of each unitary and also reduces the number of measurements required per time step,  leading to a quadratic speedup over QITE~\cite{tan2020fast}. A time dependent drifted QITE  introduces the concept of randomised compiling,  which reduces the unitary circuit depth to be a constant and  also  reduces the number of measurements needed~\cite{huang2022efficient}. We also have an implementation of  QITE using nonlocal approximation, which reduces circuit depth and is NISQ-friendly~\cite{nishi2021implementation}.

\begin{figure*}[b]
    \centering
    \includegraphics[width=\textwidth]{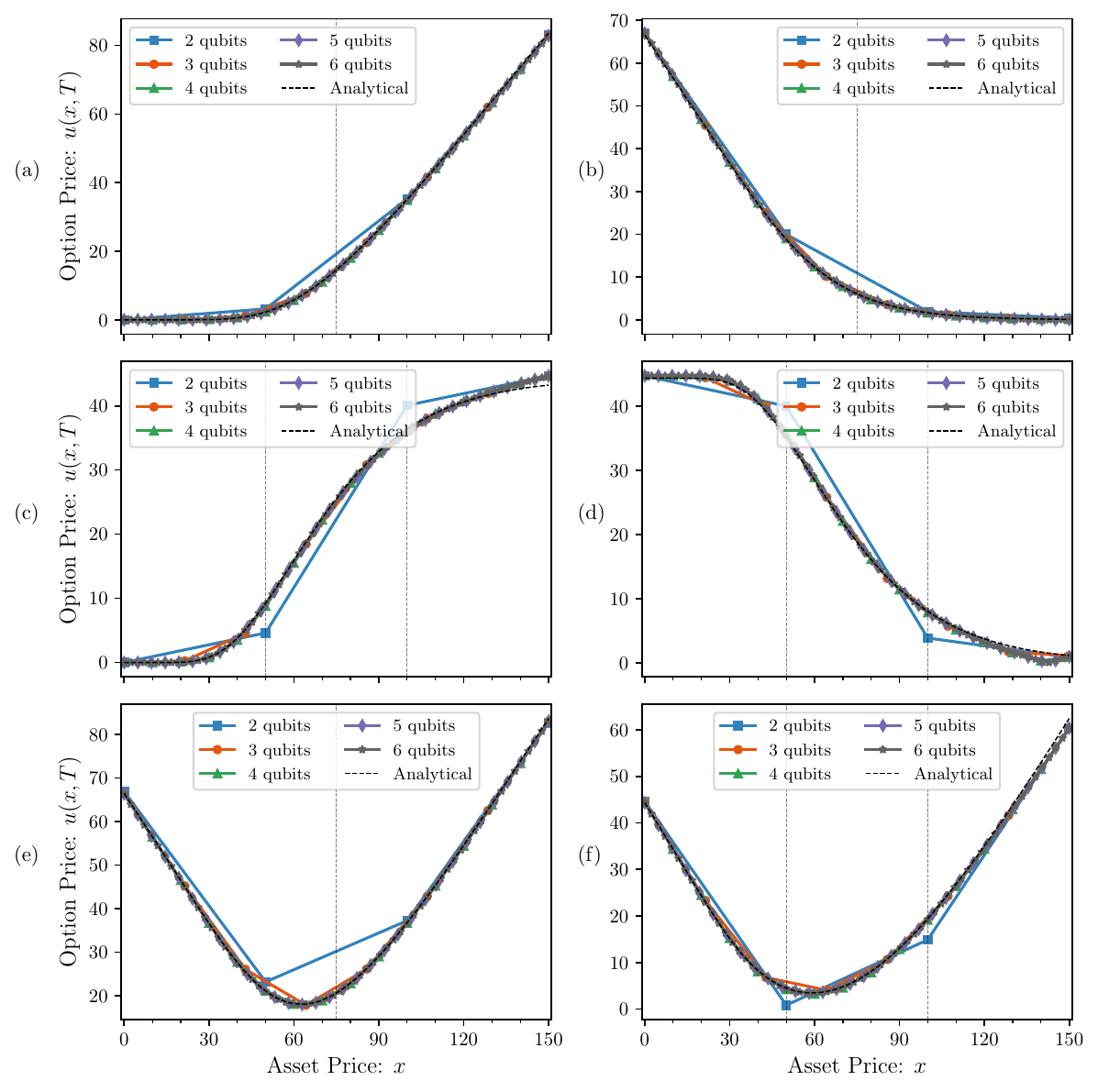}
    \caption{\textbf{Black-Scholes option pricing simulations using QNUTE.} 
    The figure compares the Black-Scholes option prices calculated using QNUTE with varying number of qubits to the corresponding analytical solutions for the following European option types:
    (a) Call (b) Put (c) Bull Spread (d) Bear Spread (e) Straddle (f) Strangle.
    The vertical dashed lines at $x=50,75,$ and $100$ correspond to the strike prices of the option contracts.
    We simulated the solutions for the asset prices $x\in[0,150]$, with the maturity time $T=3$ years, simulated over $N_T=500$ time steps. Our simulations used a risk-free interest rate of $r=0.04$, and the volatility $\sigma=0.2$. The unitaries used to approximate the evolution act on all of the qubits used in the simulation.
    }
    \label{fig:solutions}
\end{figure*}

Although imaginary time evolution was originally envisioned as a technique for determining  the ground state of a Hamiltonian~\cite{McArdle_2019, motta2020determining}, the methodology   has been recently used  as an approach for solving  partial differential equations (PDEs), primarily based on varQITE~\cite{nguyen2024solving, leong2023variational, Alghassi2022variationalquantum, fontanela2021quantum, liu2021}.
However, a simulated QITE approach for solving  linear PDEs was recently considered~\cite{kumar2024generalising}, although it is restricted to anti-Hermitian Hamiltonians involving only even-ordered derivatives. This application tracks how the non-unitary time evolution scales the quantum state over time. The approach was used to generate solutions to the isotropic heat equation by combining the scale information with the normalised states obtained from QITE.

In this paper, we further widen the scope of simulated QITE by broadening the methodology to simulations involving arbitrary non-Hermitian dynamics. By removing simulated QITE's underlying restriction to anti-Hermitian Hamiltonians, we enhance the capabilities of the methodology with an ability to simulate arbitrary linear PDEs involving non-unitary time evolution. We have called this generalisation of simulated QITE to arbitrary Hamiltonians quantum non-unitary time evolution (QNUTE).  

\section*{Results}

\noindent\textbf{Quantum Non-Unitary Time Evolution} 

\noindent QNUTE is a  quantum algorithm that simulates the dynamics of the Schr\"odinger equation with an arbitrary  non-Hermitian Hamiltonian ${\hat{H} = \sum_{m=1}^M i\hat{h}_m}$. The non-unitary time evolution operator generated by $\hat{H}$ is approximated by its first order Trotter product, and takes the form
\begin{equation}
e^{-i\hat{H}T} \approx \left(\prod_{m=1}^M e^{\hat{h}_m \Delta t}\right)^{N_T}, 
\end{equation}
where $N_T = T/\Delta t$~\cite{JonesOpt19, trotter59}.
The normalised actions of each Trotter step $e^{\hat{h}_m\Delta t}$ acting on a state $\ket{\psi}$ are approximated with unitaries  of the form $e^{-i\hat{A}\Delta t}$, and implemented with Trotter products of the form
\begin{equation}
    e^{-i\hat{A}\Delta t} \approx \prod_{I=1}^{\mathcal{I}} e^{-i a_I \hat{\sigma}_I \Delta t}.\label{eqn:tprod}
\end{equation}
In Eq.~(\ref{eqn:tprod}),  ${\hat{A}=\sum_{I=1}^{\mathcal{I}} a_I \hat{\sigma}_I}$ is a Hermitian operator with $\hat{\sigma}_I$ denoting 
Hermitian operators chosen such that each unitary $e^{-i \theta \hat{\sigma}_I }$ is efficiently implemented with a quantum circuit parameterised by $\theta$.
The real-valued coefficients $a_I$ are determined by minimising the expression
\begin{equation}
    \left\|
    \frac{e^{\hat{h}_m \Delta t}\ket{\psi}}
    {\sqrt{
    \ev{e^{\hat{h}_m^\dagger \Delta t} e^{\hat{h}_m \Delta t}}{\psi}
    }} - 
    e^{-i\hat{A}\Delta t}\ket{\psi}
    \right\|,
\end{equation}
up to $O(\Delta t)$, which  involves solving  a system of linear equations, ${(S+S^\top)\, \vec{a}=\vec{b}}$, constructed using various measurements on $\ket{\psi}$. In particular, we have   
\begin{eqnarray}    
\begin{aligned}
    S_{I,J} &= \ev{\hat{\sigma}_I^\dagger\hat{\sigma}_J}{\psi},  \\
    c &= \sqrt{1 + 2\Delta t\, \text{Re}\ev{\hat{h}_m}{\psi}},\\
        b_I &= \frac{-2}{c}\, \text{Im}\ev{\hat{\sigma}_I^\dagger\, \hat{h}_m}{\psi},  
\end{aligned}
\label{eqn:qnute-measurements}
\end{eqnarray}
see Supplementary Information for further details on the construction.
Simulating each Trotter step involves taking $O(\mathcal{I}^2)$ measurements to construct the $\mathcal{I}\times\mathcal{I}$ matrix equation and  generates a quantum circuit of depth $O(\mathcal{I})$. The full simulation therefore requires $O(N_T M \mathcal{I}^2)$ measurements.

The states generated by QNUTE are determined by the choice of $\hat{\sigma}_I$. For example, choosing $\hat{\sigma}_I$ to encompass  all Pauli strings allows us to capture arbitrary state vector rotations in the state space, whereas  restricting $\hat{\sigma}_I$ to Pauli strings involving an odd number of $\hat{Y}$ gates significantly reduces the operator decomposition count and allows us to capture those  rotations that do not introduce complex phases to the quantum state. 
Given a choice of $\hat{\sigma}_I$, the accuracy of the QNUTE implementation is dependent on the support of  $\hat{A}$.
Ideally, the support of $\hat{A}$ should cover ${D=O(C)}$ adjacent qubits surrounding the support of $\hat{h}_m$, where  the correlation length $C$ denotes the maximum distance between interacting qubits in the Hamiltonian. However, our choice to express $\hat{A}$ has been in terms of Pauli strings, which gives rise to an exponential dependence on $D$, ${\mathcal{I}=O(2^D)}$. For this reason,  we have considered an inexact implementation of QNUTE that uses a constant domain size $D<C$.

We will demonstrate that  QNUTE can be used to  approximate solutions to arbitrary linear PDEs with solutions stored  in the qubit state vector. Information relevant to the solution is extracted by taking measurements on the final quantum state. It is expected that the number of distinct measurements required to extract the relevant information should scale polynomially with the number of qubits.
Further, if it is known that the solution to a PDE will be real-valued and non-negative, then the normalised solution calculated by QNUTE can be extracted obtained by taking the square root of the probability distribution of computational basis states. We will use QNUTE to  simulate the Black-Scholes equation, as it has a non-Hermitian Hamiltonian and has non-negative real-valued solutions. 

\begin{figure*}[t]
    \centering
    \includegraphics[width=\textwidth]{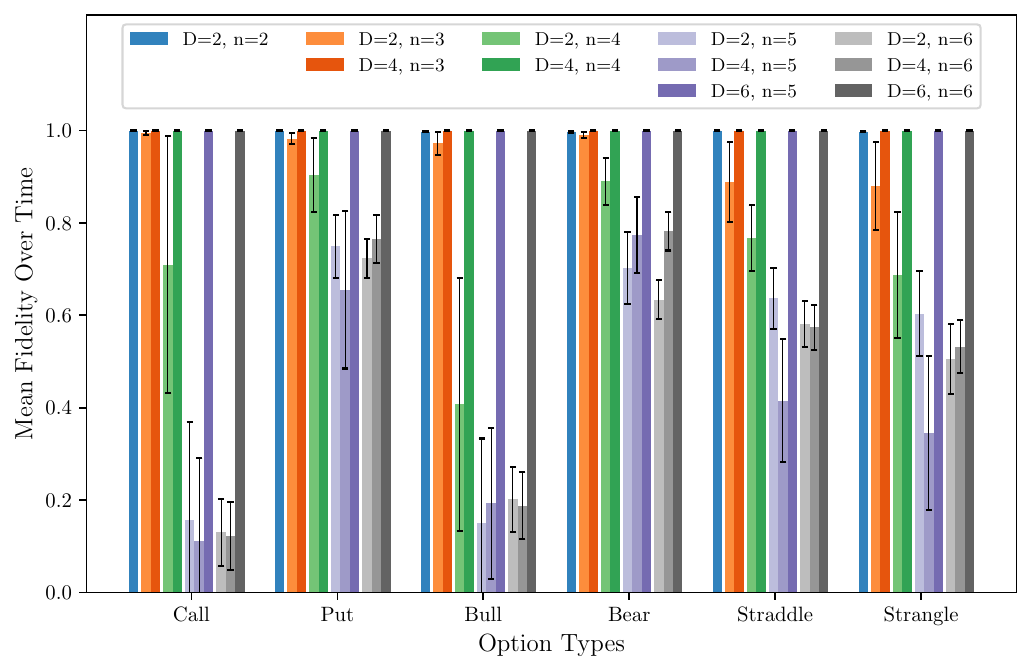}
    \caption{\textbf{Average fidelities of inexact QNUTE implementations.} The figure shows the fidelities of different implementations of inexact QNUTE used to simulate Black-Scholes dynamics averaged over each time step, with the error bars depicting the standard deviation. These simulations share the same parameters values for $r,\sigma,T$ and $N_T$ as with the simulations shown in Fig.~\ref{fig:solutions}. 
    $n$ denotes the number of qubits used to store the function samples, and $D$ denotes the maximum number of adjacent qubits targeted by the unitaries.
    The overall low fidelities shown the by inexact QNUTE, where $D<n$, indicate that the Black-Scholes Hamiltonian with linear boundary conditions has a high correlation length, making it difficult to accurately reproduce its evolution with small unitaries.}
    \label{fig:mean-fidelity}
\end{figure*}

\noindent\textbf{Simulating Black-Scholes with QNUTE}

\noindent To model the dynamics of the Black-Scholes equation, we discretise the domain $[x_0,x_N]$ into $2^n$ points equally spaced by a distance of $h = \frac{x_N - x_0}{2^n - 1}$.  The normalised samples of the option price are stored in an $n$-qubit quantum state given by 
\begin{equation}
    \ket{\bar{u}(\tau)} = \frac{\sum_{k=0}^{2^n-1} u(x_k, \tau) \ket{k}}
    {\sqrt{\sum_{k=0}^{2^n-1} u^2(x_k, \tau) }}, 
    \label{eqn:encoding-scheme}
\end{equation}
where $x_k = x_0 + kh$. Following Eq.~(\ref{eqn:Hbs}), the discretised Black-Scholes Hamiltonian can be represented  in terms of a central finite  difference matrix of the form 
\begin{equation}
    -i\hat{H}_{BS} = 
    \begin{bmatrix}
        \gamma_0 & \beta_0 &  &  \\
        \alpha_1 & \gamma_1 & \beta_1 & \\
        &  \ddots& \ddots & \ddots & &\\
        & & \alpha_{2^n-2} & \gamma_{2^n-2} & \beta_{2^n-2} \\
        & & & \alpha_{2^n-1} & \gamma_{2^n-1}
    \end{bmatrix},
    \label{eqn:Hbs-central}
\end{equation}
where
\begin{eqnarray}    
\begin{aligned}
    \alpha_k &= \frac{\sigma^2 x_k^2}{2h^2} - \frac{r x_k}{2h}, \\
    \beta_k &= \frac{\sigma^2 x_k^2}{2h^2} + \frac{r x_k}{2h}, \\
    \gamma_k &= -r - \alpha_k - \beta_k.
\end{aligned}
\end{eqnarray}
Refer to Supplementary Information for the representation of the discretised Hamiltonian of Eq.~(\ref{eqn:Hbs-central}) in the Pauli operator basis.

\noindent\textbf{Norm correction}

\noindent
The scale factor $c$ given in Eq.~(\ref{eqn:qnute-measurements})  approximates  how the Trotter step scales $\ket{\psi}$
up to $O(\Delta t)$. 
These approximations can be stored and multiplied to provide an approximation of how the state vector scales over the course of the evolution. Excluding the scenario of the  ideal implementation of QNUTE  that  records a perfect fidelity, errors associated to each scale factor will compound   over multiple Trotter steps, which must be corrected periodically. 
For an anti-Hermitian Hamiltonian $\hat{H}=i\hat{L}$, it was shown that the correction factor can be calculated using knowledge of the non-degenerate ground state $\ket{\psi_0}$ of $\hat{L}$ and its corresponding eigenvalue $\lambda_0$~\cite{kumar2024generalising}. This correction strategy necessarily exploits the mutual orthogonality of the eigenstates of the associated Hamiltonian.

However, since the discretised Black-Scholes Hamiltonian as given in Eq.~(\ref{eqn:Hbs-central}) is not a normal operator, its eigenvectors are not  guaranteed to be mutually orthogonal.  This, therefore, rules out the norm correction strategy pursued in Ref.~\cite{kumar2024generalising}.  Interestingly,   variational QITE  has been employed as a technique for solving the Black-Scholes equation. Under this setting, the normalisation factor was considered either as a variational parameter~\cite{Alghassi2022variationalquantum} or was determined with prior knowledge of how, specifically, call option prices evolve at the boundary $x_N$~\cite{fontanela2021quantum}. Since the former is not compatible with QNUTE, we generalise the latter approach to cater to various European option types.

\begin{table}[!htbp]
\caption{Average fidelity data for the QNUTE simulations for pricing various European option types. $\mu_F$  denotes the mean fidelity over each time step and $\sigma_F$  the standard deviation. Simulations involving $D<n$ represent inexact QNUTE implementations.}
\centering
\begin{tabular}{ll ll ll}
\toprule
\multirow{2}{*}{$n$} & \multirow{2}{*}{$D$} & \multicolumn{2}{c}{Call} & \multicolumn{2}{c}{Put}\\ \cmidrule(lr){3-4} \cmidrule(lr){5-6}
 & & \multicolumn{1}{c}{$\mu_F$} & \multicolumn{1}{c}{$\sigma_F$} & \multicolumn{1}{c}{$\mu_F$} & \multicolumn{1}{c}{$\sigma_F$} \\ \toprule
2 & 2 & 1.000 & 2.67$\times 10^{-5}$ & 1.000 & 4.54$\times 10^{-5}$ \\ \midrule
\multirow[c]{2}{*}[-0.3em]{3} & 2& 0.994 & 4.77$\times 10^{-3}$ & 0.982 & 1.20$\times 10^{-2}$ \\ \cmidrule(l){2-6}
 & 4& 1.000 & 3.26$\times 10^{-6}$ & 1.000 & 3.06$\times 10^{-6}$ \\ \midrule
\multirow[c]{2}{*}[-0.3em]{4} & 2& 0.710 & 2.78$\times 10^{-1}$ & 0.904 & 8.02$\times 10^{-2}$ \\ \cmidrule(l){2-6}
 & 4& 1.000 & 3.16$\times 10^{-7}$ & 1.000 & 2.79$\times 10^{-7}$ \\ \midrule
\multirow[c]{3}{*}[-0.3em]{5} & 2& 0.158 & 2.11$\times 10^{-1}$ & 0.749 & 6.79$\times 10^{-2}$ \\ \cmidrule(l){2-6}
 & 4& 0.111 & 1.79$\times 10^{-1}$ & 0.655 & 1.70$\times 10^{-1}$ \\ \cmidrule(l){2-6}
 & 6& 1.000 & 4.15$\times 10^{-8}$ & 1.000 & 2.10$\times 10^{-7}$ \\ \midrule
\multirow[c]{3}{*}[-0.6em]{6} & 2& 0.130 & 7.27$\times 10^{-2}$ & 0.724 & 4.18$\times 10^{-2}$ \\ \cmidrule(l){2-6}
 & 4& 0.122 & 7.32$\times 10^{-2}$ & 0.766 & 5.20$\times 10^{-2}$ \\ \cmidrule(l){2-6}
 & 6& 1.000 & 3.81$\times 10^{-8}$ & 1.000 & 2.18$\times 10^{-7}$ \\ \bottomrule
\end{tabular}
\vspace{1pt}
\begin{tabular}{ll ll ll}
\toprule
\multirow{2}{*}{$n$} & \multirow{2}{*}{$D$} & \multicolumn{2}{c}{Bull} & \multicolumn{2}{c}{Bear}\\ \cmidrule(lr){3-4} \cmidrule(lr){5-6}
 & & \multicolumn{1}{c}{$\mu_F$} & \multicolumn{1}{c}{$\sigma_F$} & \multicolumn{1}{c}{$\mu_F$} & \multicolumn{1}{c}{$\sigma_F$} \\ \toprule
2 & 2 & 1.000 & 2.67$\times 10^{-5}$ & 1.000 & 4.54$\times 10^{-5}$ \\ \midrule
\multirow[c]{2}{*}[-0.3em]{3} & 2& 0.994 & 4.77$\times 10^{-3}$ & 0.982 & 1.20$\times 10^{-2}$ \\ \cmidrule(l){2-6}
 & 4& 1.000 & 3.26$\times 10^{-6}$ & 1.000 & 3.06$\times 10^{-6}$ \\ \midrule
\multirow[c]{2}{*}[-0.3em]{4} & 2& 0.710 & 2.78$\times 10^{-1}$ & 0.904 & 8.02$\times 10^{-2}$ \\ \cmidrule(l){2-6}
 & 4& 1.000 & 3.16$\times 10^{-7}$ & 1.000 & 2.79$\times 10^{-7}$ \\ \midrule
\multirow[c]{3}{*}[-0.3em]{5} & 2& 0.158 & 2.11$\times 10^{-1}$ & 0.749 & 6.79$\times 10^{-2}$ \\ \cmidrule(l){2-6}
 & 4& 0.111 & 1.79$\times 10^{-1}$ & 0.655 & 1.70$\times 10^{-1}$ \\ \cmidrule(l){2-6}
 & 6& 1.000 & 4.15$\times 10^{-8}$ & 1.000 & 2.10$\times 10^{-7}$ \\ \midrule
\multirow[c]{3}{*}[-0.6em]{6} & 2& 0.130 & 7.27$\times 10^{-2}$ & 0.724 & 4.18$\times 10^{-2}$ \\ \cmidrule(l){2-6}
 & 4& 0.122 & 7.32$\times 10^{-2}$ & 0.766 & 5.20$\times 10^{-2}$ \\ \cmidrule(l){2-6}
 & 6& 1.000 & 3.81$\times 10^{-8}$ & 1.000 & 2.18$\times 10^{-7}$ \\ \bottomrule
\end{tabular}
\vspace{1pt}
\begin{tabular}{ll ll ll}
\toprule
\multirow{2}{*}{$n$} & \multirow{2}{*}{$D$} & \multicolumn{2}{c}{Straddle} & \multicolumn{2}{c}{Strangle}\\ \cmidrule(lr){3-4} \cmidrule(lr){5-6}
 & & \multicolumn{1}{c}{$\mu_F$} & \multicolumn{1}{c}{$\sigma_F$} & \multicolumn{1}{c}{$\mu_F$} & \multicolumn{1}{c}{$\sigma_F$} \\ \toprule
2 & 2 & 1.000 & 2.67$\times 10^{-5}$ & 1.000 & 4.54$\times 10^{-5}$ \\ \midrule
\multirow[c]{2}{*}[-0.3em]{3} & 2& 0.994 & 4.77$\times 10^{-3}$ & 0.982 & 1.20$\times 10^{-2}$ \\ \cmidrule(l){2-6}
 & 4& 1.000 & 3.26$\times 10^{-6}$ & 1.000 & 3.06$\times 10^{-6}$ \\ \midrule
\multirow[c]{2}{*}[-0.3em]{4} & 2& 0.710 & 2.78$\times 10^{-1}$ & 0.904 & 8.02$\times 10^{-2}$ \\ \cmidrule(l){2-6}
 & 4& 1.000 & 3.16$\times 10^{-7}$ & 1.000 & 2.79$\times 10^{-7}$ \\ \midrule
\multirow[c]{3}{*}[-0.3em]{5} & 2& 0.158 & 2.11$\times 10^{-1}$ & 0.749 & 6.79$\times 10^{-2}$ \\ \cmidrule(l){2-6}
 & 4& 0.111 & 1.79$\times 10^{-1}$ & 0.655 & 1.70$\times 10^{-1}$ \\ \cmidrule(l){2-6}
 & 6& 1.000 & 4.15$\times 10^{-8}$ & 1.000 & 2.10$\times 10^{-7}$ \\ \midrule
\multirow[c]{3}{*}[-0.6em]{6} & 2& 0.130 & 7.27$\times 10^{-2}$ & 0.724 & 4.18$\times 10^{-2}$ \\ \cmidrule(l){2-6}
 & 4& 0.122 & 7.32$\times 10^{-2}$ & 0.766 & 5.20$\times 10^{-2}$ \\ \cmidrule(l){2-6}
 & 6& 1.000 & 3.81$\times 10^{-8}$ & 1.000 & 2.18$\times 10^{-7}$ \\ \bottomrule
\end{tabular}
\label{tab:fidelities}
\end{table}

Consider the Black-Scholes equation, as  given in Eq.~(\ref{eqn:Hbs}), 
with option price 
$u(x,\tau)$ assumed  to be linear in $x$ in the neighbourhood of the boundaries $x_0$ and $x_N$.
We will consider linear boundary conditions, since they are widely used in classical option pricing simulations and are known to be numerically stable~\cite{Windcliff2004AnalysisOT}. 
Thus, under linear boundary conditions, the option price takes the form ${u(x,\tau) = a(\tau)x + b(\tau)}$ near the boundaries. Substituting this form into Eq.~(\ref{eqn:Hbs}) reduces the Black-Scholes equation to an ordinary differential equation (ODE) at the boundaries
\begin{equation}
    x\dv{a}{\tau} + \dv{b}{\tau} = -rb(\tau).
    \label{eqn:ode}
\end{equation}
Solving Eq.~(\ref{eqn:ode}) yields  $a(\tau) = a(0)$ and $b(\tau) = b(0)e^{-r\tau}$, where $a(0)$, and $b(0)$ can be derived from the initial conditions $p(x)$ at each boundary. If $a(0)$ or $b(0)$ are non-zero on at least one of the boundaries, we can rescale a normalised solution to ensure that the value at that boundary is equal to $a(\tau)x+b(\tau)$.

To guarantee that the linear boundary conditions apply during the QNUTE simulation, they must be encoded into the Black-Scholes Hamiltonian. The first and last rows of the matrix in Eq.~(\ref{eqn:Hbs-central}) are updated with the corresponding forward and backward first-order finite difference coefficients, respectively, with the second-derivative terms vanishing as the function is linear. The Black-Scholes Hamiltonian inclusive of linear boundary conditions takes the form 
\begin{equation}
    -i\hat{H}_{LBS} = \begin{bmatrix}
        \gamma_0^\prime & \beta_0^\prime &  &  \\
        \alpha_1 & \gamma_1 & \beta_1 & \\
        &  \ddots& \ddots & \ddots & &\\
        & & \alpha_{2^n-2} & \gamma_{2^n-2} & \beta_{2^n-2} \\
        & & & \alpha_{2^n-1}^\prime & \gamma_{2^n-1}^\prime
    \end{bmatrix},
\end{equation}
where
\begin{eqnarray}
\begin{aligned}
    \gamma_0^\prime       &=\ -r - \frac{r x_0}{h}, \\
    \beta_0^\prime        &=\ \frac{r x_0}{h}, \\
    \alpha_{2^n-1}^\prime &=\ -\frac{r x_N}{h}, \\
    \gamma_{2^n-1}^\prime &=\ -r + \frac{r x_N}{h}.
\end{aligned}
\end{eqnarray}
See Supplementary Information for the Pauli decomposition of this Hamiltonian.

\section*{Discussion}
In this work, we have generalised the quantum imaginary time evolution algorithm to enable the simulation of arbitrary non-Hermitian dynamics on quantum computers. We demonstrated our QNUTE algorithm's application via a numerical implementation that simulating the pricing dynamics of European options, as dictated by the Black-Scholes equation.

In undertaking these simulations, we assumed that the underlying financial asset had constant volatility, $\sigma$,  and risk-free interest rate, $r$. The time dependence of these variables can be encoded in the Hamiltonian with no extra cost to its construction. Further, the inclusion of these variables does not affect the unitary approximations produced by QNUTE, however, modelling volatility and interest rates as stochastic processes may require smaller time steps for more accurate simulations. Indeed, the time dependence of $r$ gives rise to a different boundary ODE, which necessitates modifications to our rescaling protocol.

As depicted in Fig.~\ref{fig:solutions}, our implementations of QNUTE were able to match the analytical solutions of the Black-Scholes equation. For  convergence, it is important to choose an asset price domain with boundaries such that linear boundary conditions hold for the option's payoff $u(x,\tau=T)$. A good level of convergence also depends on having access to enough sample points around the strike prices of the option, a lack of which can be seen in the 2-qubit curves in Fig.~\ref{fig:solutions} (c), (d) and (f).

In our implementations of QNUTE, each term, $\hat{h}_m$, in the decomposition of the Black-Scholes Hamiltonian was a linear combination of Pauli strings. Since the number of distinct Pauli strings in our  decomposition scales exponentially with the number of qubits, an alternative decomposition is required for the scalability of QNUTE for Black-Scholes. 
Approaches taken to solve PDEs using varQITE have also required the expectation values of finite difference operators. In particular, Liu et al. proposed a scheme to measure such expectation values with a linear overhead~\cite{liu2021}. We conjecture that this scheme may be adopted within our approach, leading to exponentially fewer terms in the Hamiltonian decomposition.

The discretised Black-Scholes Hamiltonian has a high degree of correlation between all the qubits used in the simulation. This was demonstrated  by implementing inexact QNUTE, wherein the unitary approximations only act on at most $D$ adjacent qubits. Figure~\ref{fig:mean-fidelity} depicts the fidelities of inexact QNUTE simulations, averaged over each time step for the various option payoffs, see Table~\ref{tab:fidelities} for the exact values. For an increasing number of qubits and fixed domain size $D$, we have it that  inexact QNUTE does not capture the correlations between the qubits, rendering it unable to emulate the true time evolution.
For future work, we intend to incorporate  recent improvements to the simulated QITE methodology, including Fast QITE~ \cite{tan2020fast}, time-dependent drifted QITE~\cite{huang2022efficient} and QITE with nonlocal approximation~\cite{nishi2021implementation},  within our QNUTE framework to understand their effect on the accuracy and efficiency for simulated PDE dynamics.

\section*{Data availability}
All data generated and analysed during this study are included in this published article and its supplementary information files. 

\section*{Code availability}
The code that supports the findings of this study is  available from the corresponding authors upon reasonable request.

\bibliography{sn-bibliography} 

\section*{Acknowledgements}
C.M.W. and S.K. gratefully acknowledge the financial support of the Engineering and Physical Sciences Research Council (EPSRC) through the Hub in Quantum Computing and Simulation (EP/T001062/1).

\section*{Author Contributions}
C.M.W. conceived the project. S.K. performed the numerical simulations. C.M.W. and S.K. analyzed the data and interpreted the results, and proposed improvements. C.M.W. and S.K. both contributed to the writing and editing of the manuscript.

\section*{Competing Interests}
The authors declare no competing interests.

\onecolumn
\section*{Supplementary Information}
\subsection*{Calculating the unitaries} 
We approximate the normalised action of a Trotter step $e^{\hat{h}_m \Delta t}$ on the state $\ket{\psi}$ by using the  unitary $e^{-i\hat{A}\Delta t}$, where  $\hat{A} = \sum_I a_I \hat{\sigma_I}$ is a Hermitian operator expressed as a real linear combination of Pauli operator strings indexed by $I$. To calculate the coefficients $a_I$, we minimise the distance between the normalised Trotter step and the unitary acting on $\ket{\psi}$ expanded to $O(\Delta t)$. This is achieved by firstly approximating  the norm of the Trotter step acting on $\ket{\psi}$. We have it that   
\begin{eqnarray}\begin{aligned}
    c^2 &= {\ev{e^{\hat{h}_m^\dagger\Delta t} e^{\hat{h}_m \Delta t}}{\psi}} \\
    &= \ev{(\hat{\mathbbm{1}} + \hat{h}_m^\dagger \Delta t)(\hat{\mathbbm{1}} + \hat{h}_m \Delta t)}{\psi} + O(\Delta t^2)   \\
    &= \ev{\hat{\mathbbm{1}} + (\hat{h}_m^\dagger + \hat{h}_m) \Delta t}{\psi} +O(\Delta t^2)  \\
    &= 1 + 2\mathrm{Re}\ev{\hat{h}_m}{\psi}\Delta t + O(\Delta t^2),   
\end{aligned}
\end{eqnarray}
from which we have 
\begin{eqnarray}
     c &\approx \sqrt{1 + 2\Delta t\,\mathrm{Re}\ev{\hat{h}_m}{\psi}}.
\end{eqnarray}
Next, we set the unitary $e^{-i\hat{A}\Delta t}$ to map $\ket{\psi}$ to  the normalised output of the Trotter step
\begin{eqnarray}
    e^{-i\hat{A}\Delta t}\ket{\psi} &= \frac{e^{\hat{h}_m\Delta t}}{c}\ket{\psi}, 
    \end{eqnarray}
    from which we have
    \begin{align}
      (\hat{\mathbbm{1}} - i\hat{A}\Delta t)\ket{\psi}
    &\approx \left(\frac{\hat{\mathbbm{1}} + \hat{h}_m\Delta t}{c}\right) \ket{\psi}, \end{align}
   and rewriting  
   \begin{align}
      -i\hat{A}\ket{\psi} 
    &\approx \frac{1-c}{c\,\Delta t}\ket{\psi} 
    + \frac{\hat{h}_m}{c}\ket{\psi}.
\end{align}
Now, defining
\begin{equation}
    \ket{\Delta_0}:=\frac{1-c}{c\,\Delta t}\ket{\psi} 
    + \frac{\hat{h}_m}{c}\ket{\psi},
\end{equation}
and
\begin{equation}
    \ket{\Delta} := -i\hat{A}\ket{\psi}, 
\end{equation}
we solve for $\hat{A}$ such that  the squared distance between $\ket{\Delta}$ and $\ket{\Delta_0}$ is minimised. It follows that 
\begin{eqnarray}\begin{aligned}
    \| \ket{\Delta_0} - \ket{\Delta} \|^2 &= 
    (\bra{\Delta_0} - \bra{\Delta}) (\ket{\Delta_0} - \ket{\Delta})  \\
    &= \braket{\Delta_0} - \braket{\Delta_0}{\Delta} - \braket{\Delta}{\Delta_0} + \braket{\Delta}  \\
    &= \braket{\Delta_0} + \braket{\Delta} - 2\mathrm{Re}\braket{\Delta_0}{\Delta}.
\end{aligned}
\end{eqnarray}
Since $\ket{\Delta_0}$ does not depend on $\hat{A}$, $\braket{\Delta_0}$, in the minimisation,  can be ignored. Noting that $\hat{A}$ is Hermitian, we see that
\begin{equation}
    \braket{\Delta} = \ev{\hat{A}^2}{\psi},
\end{equation}
while the quantity  $\braket{\Delta_0}{\Delta}$ can be evaluated as 
\begin{eqnarray}
     \braket{\Delta_0}{\Delta}
    &=& \left(\frac{1-c}{c\,\Delta t}\bra{\psi} + \bra{\psi}\frac{\hat{h}_m^\dagger}{c}\right)
    (-i\hat{A}\ket{\psi})
      \\
    &=& -i\left(\frac{1-c}{c\,\Delta t}\right)\ev{\hat{A}}{\psi}
    -\frac{i}{c}\ev{\hat{h}_m^\dagger \hat{A}}{\psi}.\label{2ndRE}
 \end{eqnarray}
Using that $\hat{A}$ is Hermitian, the expectation value of $\ev{\hat{A}}{\psi}$ will always be   real, which implies that the real part of $\braket{\Delta_0}{\Delta}$ only depends on the second term in Eq.~(\ref{2ndRE})
\begin{eqnarray}\begin{aligned}
    \mathrm{Re}(\braket{\Delta_0}{\Delta}) 
    &= \mathrm{Re}\left( \frac{-i}{c} \ev{\hat{h}_m^\dagger \hat{A}}{\psi} \right)
     \\
    &= \mathrm{Im}\left(\frac{\ev{\hat{h}_m^\dagger \hat{A}}{\psi}}{c}\right)
      \\
    &= -\mathrm{Im}\left(\frac{\ev{\hat{A}\hat{h}_m}{\psi}}{c}\right).
\end{aligned}\end{eqnarray}
The  function to minimise, therefore, reduces to 
\begin{eqnarray}\begin{aligned}
    f(\vec{a}) 
    &= -2\mathrm{Re}(\braket{\Delta_0}{\Delta} + \braket{\Delta})
      \\
    &= \frac{2}{c}\,\mathrm{Im}(\ev{\hat{A}\hat{h}_m}{\psi}) + \ev{\hat{A}^2}{\psi} 
  \\
    &=\sum_I \frac{2a_I}{c}\, \mathrm{Im}(\ev{\hat{\sigma}_I \hat{h}_m}{\psi}) 
    + \sum_{I,J} a_I a_J \ev{\hat{\sigma}_I\hat{\sigma}_J}{\psi}.
\end{aligned}\end{eqnarray}
Defining   $\vec{b}$ with components
\begin{equation}
    b_I := \frac{-2}{c}\,\mathrm{Im}(\ev{\hat{\sigma}_I\hat{h}_m}{\psi}),
\end{equation}
and the matrix $S$ with entries
\begin{equation}
    S_{I,J} := \ev{\hat{\sigma}_I \hat{\sigma}_J}{\psi},
\end{equation}
we can rewrite the objective function as
\begin{eqnarray}\begin{aligned}
    f(\vec{a}) 
    &= \sum_{I,J} a_I S_{I,J} a_J - \sum_{I} b_I a_I
    \\
    &= \vec{a} \cdot S \vec{a} - \vec{b} \cdot \vec{a}.
\end{aligned}\end{eqnarray}
To minimise $f$ with respect to $\vec{a}$, we equate its gradient to zero, and observe that \begin{eqnarray}\begin{aligned}
    \pdv{f}{a_I} 
    &= \sum_{J\neq I} (S_{I,J} + S_{J,I}) a_J + 2 S_{I,I} a_I - b_I
     \\
    &= \sum_{I,J} (S_{I,J} + S_{J,I}) a_J - b_I. 
     \end{aligned}\end{eqnarray}
It follows that $\nabla f = (S + S^\top)\vec{a} - \vec{b}$, and, hence,  to minimise $f$, we solve for $\vec{a}$ in
\begin{equation}
    (S+S^\top)\vec{a} = \vec{b}.
\end{equation}

\subsection*{The Black-Scholes Hamiltonian in the Pauli basis}
In order to discretise the Black-Scholes equation with the form, as given in Eq.~(\ref{eqn:black-scholes-eqn}),  
\begin{equation}
    \pdv{u}{\tau} = \frac{\sigma^2}{2} x^2 \pdv[2]{u}{x} + r x\pdv{u}{x} - ru,
    \label{SI-eqn:black-scholes}
\end{equation}
we rewrite the derivative operators in terms of   finite difference representations and substitute the $x$ and $x^2$ terms with the diagonal matrices
\begin{equation}
    {\hat{X} = \begin{bmatrix}
        x_0 & & \\
          & \ddots & \\
       & & x_N   
    \end{bmatrix}} \ \textrm{and} \ 
    {\hat{X}^2 = \begin{bmatrix}
        x_0^2 & & \\
          & \ddots & \\
       & & x_N^2   
    \end{bmatrix}},
\end{equation}
respectively, where $x_k = x_0 + kh$.
We are required to represent   these matrices as linear combinations of the Pauli operators in order to be able to simulate their dynamics with QNUTE. Firstly, let us define the following one-qubit matrices.  
\begin{eqnarray}\begin{aligned}
     \hat{A}_\nwarrow^{(1)} := \begin{bmatrix}
         1&0\\
         0&0
     \end{bmatrix}
     &= 
     \frac{\hat{I} + \hat{Z}}{2}
     \\
     \hat{A}_\searrow^{(1)} := \begin{bmatrix}
         0&0\\
         0&1
     \end{bmatrix}
     &= \frac{\hat{I} - \hat{Z}}{2}
     \\
     \hat{A}_\nearrow^{(1)} := \begin{bmatrix}
         0&1\\
         0&0
     \end{bmatrix}
     &= \frac{\hat{X} + i\hat{Y}}{2}
     \\
     \hat{A}_\swarrow^{(1)} := \begin{bmatrix}
         0&0\\
         1&0
     \end{bmatrix}
     &= \frac{\hat{X} - i\hat{Y}}{2}
\end{aligned}\label{eqn:1-q}
\end{eqnarray}
Further, consider the $n$-qubit generalisations of Eq.~(\ref{eqn:1-q}) given by
\begin{eqnarray}\begin{aligned}
    \hat{A}_\nwarrow^{(n)} &:= \hat{A}_\nwarrow^{(1)} \otimes \hat{A}_\nwarrow^{(n-1)},
     \\
     \hat{A}_\searrow^{(n)} &:= \hat{A}_\searrow^{(1)} \otimes \hat{A}_\searrow^{(n-1)},
     \\
     \hat{A}_\nearrow^{(n)} &:= \hat{A}_\nearrow^{(1)} \otimes \hat{A}_\nearrow^{(n-1)},
     \\
     \hat{A}_\swarrow^{(n)} &:= \hat{A}_\swarrow^{(1)} \otimes \hat{A}_\swarrow^{(n-1)}.
\end{aligned}\end{eqnarray}
Tensor product strings comprising $n$ copies of the four 1-qubit operators in Eq.~(\ref{eqn:1-q}) allows us to generate matrices that have a 1 in exactly one entry of a  $2^n\times2^n$ matrix with 0s elsewhere. We will use these $n$-qubit tensor strings of the   operators in Eq.~(\ref{eqn:1-q}) to define the linear boundary conditions for  the Black-Scholes Hamiltonian.

Let us construct the $\hat{X}$ matrix for $n$-qubits. We have 
\begin{equation}
    \hat{X}^{(n)}
    = \begin{bmatrix}
        x_0 & & & \\
         & x_1 & & \\
         & & \ddots & \\
         & & & x_N
    \end{bmatrix}
    = x_0\hat{\mathbbm{1}} + h\begin{bmatrix}
        0 & & & \\
         & 1 & & \\
         & & \ddots & \\
         & & & 2^n-1
    \end{bmatrix} 
    = x_0\hat{\mathbbm{1}} + h\hat{\chi}^{(n)}.
\end{equation}
Consider the matrix $\hat{\chi}^{(n)}$.  For $n=1$,  $\hat{\chi}^{(1)} = \hat{A}_\searrow^{(1)}$, while for $n=2$, we have
\begin{equation}
    \hat{\chi}^{(2)} = \begin{bmatrix}
        \hat{\chi}^{(1)} & 0 \\ 
        0 & 2\hat{I} + \hat{\chi}^{(1)} 
    \end{bmatrix}
    = \hat{A}_\nwarrow^{(1)}\otimes\hat{\chi}^{(1)} 
    + \hat{A}_\searrow^{(1)}\otimes(2\hat{I}+\hat{\chi}^{(1)}).
\end{equation}
Generalising to $n$-qubits, we have it that 
\begin{equation}
    \hat{\chi}^{(n)} = \begin{bmatrix}
        \hat{\chi}^{(n-1)} & 0 \\ 
        0 & 2^{n-1}\hat{I}^{\otimes n-1} + \hat{\chi}^{(n-1)} 
    \end{bmatrix}
    = \hat{I}\otimes\hat{\chi}^{(n-1)} 
    + 2^{n-1} \left(\hat{A}_\searrow^{(1)}\otimes\hat{I}^{\otimes n-1}\right).
\end{equation}
Squaring both sides yields 
\begin{align}
    (\hat{\chi}^{(n)})^2 &= 
    \left(\hat{I}\otimes\hat{\chi}^{(n-1)} 
    + 2^{n-1} \left(\hat{A}_\searrow^{(1)}\otimes\hat{I}^{\otimes n-1}\right)\right)
    \left(\hat{I}\otimes\hat{\chi}^{(n-1)} 
    + 2^{n-1} \left(\hat{A}_\searrow^{(1)}\otimes\hat{I}^{\otimes n-1}\right)\right)
    \nonumber \\
    &= \hat{I}\otimes (\hat{\chi}^{(n-1)})^2 
    + 2^{n} \left(\hat{A}^{(1)}_\searrow \otimes \hat{\chi}^{(n-1)}\right)
    + 2^{2(n-1)} \left((\hat{A}^{(1)}_\searrow)^2 \otimes \hat{I}^{\otimes n-1}\right) \nonumber \\
    &= \hat{I}\otimes (\hat{\chi}^{(n-1)})^2 
    + \hat{A}^{(1)}_\searrow \otimes \left( 2^n \hat{\chi}^{(n-1)} 
    + 2^{2(n-1)} \hat{I}^{\otimes n-1} \right).
\end{align}
Next, let us consider the first derivative $\pdv{}{x}$. This operator is approximated by the central difference matrix $\frac{1}{2h}\hat{D}_1^{(n)}$, where
\begin{equation}
    \hat{D}_1^{(n)} = \begin{bmatrix}
        0 & 1 & & &\\
        -1 & 0 & 1 & &\\
         & \ddots & \ddots & \ddots &\\
         & & -1 & 0 & 1 \\
         & & & -1 & 0
    \end{bmatrix}.
\end{equation}
For one qubit case, $n=1$, we have 
\begin{equation}
    \hat{D}_1^{(1)} = \begin{bmatrix}
        0 & 1 \\
        -1 & 0 
    \end{bmatrix} = i\hat{Y}.
\end{equation}
Generalising to $n$-qubits, we have
\begin{equation}
    \hat{D}_1^{(n)} = 
    \begin{bmatrix}
         \hat{D}_1^{(n-1)} & \hat{A}_\swarrow^{(n-1)} \\ 
         -\hat{A}_\nearrow^{(n-1)} & \hat{D}_1^{(n-1)}
    \end{bmatrix}
    = \hat{I}\otimes \hat{D}_1^{(n-1)}
    + \hat{A}_\nearrow^{(1)}\otimes \hat{A}_\swarrow^{(n-1)} 
    - \hat{A}_\swarrow^{(1)}\otimes \hat{A}_\nearrow^{(n-1)}.
\end{equation}
Next, let us consider the second derivative $\pdv[2]{}{x}$, which  is approximated by  the central difference matrix $\frac{1}{h^2} \hat{D}_2^{(n)}$, where 
\begin{equation}
    \hat{D}_2^{(n)} = \begin{bmatrix}
        -2 & 1 & \\
        1 & -2 & 1 & \\
         & \ddots & \ddots & \ddots & \\
         & & 1 & -2 & 1 & \\
         & & & 1 & -2
    \end{bmatrix}.
\end{equation}
For $n=1$, we have 
\begin{equation}
    \hat{D}_2^{(1)} = \begin{bmatrix}
        -2 & 1 \\
        1 & -2
    \end{bmatrix} = -2\hat{I} + \hat{X}.
\end{equation}
Generalising to $n$-qubits, it can be shown that 
\begin{equation}
    \hat{D}_2^{(n)} = \begin{bmatrix}
        \hat{D}_2^{(n-1)} & \hat{A}_\swarrow^{(n-1)} \\
        \hat{A}_\nearrow^{(n-1)} & \hat{D}_2^{(n-1)}
    \end{bmatrix}
    = \hat{I}\otimes \hat{D}_2^{(n-1)}
    + \hat{A}_\nearrow^{(1)}\otimes \hat{A}_\swarrow^{(n-1)} 
    + \hat{A}_\swarrow^{(1)}\otimes \hat{A}_\nearrow^{(n-1)}.
\end{equation}
Using the $\hat{X}^{(n)}$, $\hat{D}_2^{(n)}$ and $\hat{D}_2^{(n)}$ representations above, the $n$-qubit Black-Scholes Hamiltonian takes the form    
\begin{equation}
    -i\hat{H}_{BS}^{(n)} = \frac{\sigma^2}{2h^2} (\hat{X}^{(n)})^2 \hat{D}_2^{(n)}
    + \frac{r}{2h}\hat{X}^{(n)}\hat{D}_1^{(n)} - r\hat{I}^{\otimes n}.\label{eqn:BSM}
\end{equation}
Expressing Eq.~(\ref{eqn:BSM}) in matrix form over $n$-qubits, we have it that   
\begin{equation}
    -i\hat{H}_{BS}^{(n)} = \begin{bmatrix}
        \gamma_0 & \beta_0 & \\
        \alpha_1 & \gamma_1 & \beta_1 & \\
        & \ddots & \ddots & \ddots & \\
        & & \alpha_{2^n-2} & \gamma_{2^n-2} & \beta_{2^n-2} \\
        & & & \alpha_{2^n-1} & \gamma_{2^n-1}
    \end{bmatrix},
\end{equation}
where 
\begin{equation}
    \alpha_k = \frac{\sigma^2 x_k^2}{2h^2} - \frac{r x_k}{2h},
    \quad \beta_k = \frac{\sigma^2 x_k^2}{2h^2} + \frac{r x_k}{2h},
    \quad\text{and}\quad\gamma_k = -r - \alpha_k - \beta_k.
\end{equation}
To include linear boundary conditions, we neglect the second derivative terms and substitute the first and last row of Eq.~(\ref{eqn:BSM}) with forward and backward finite difference coefficients, respectively. The revised first row coefficients that take into account the linear boundary conditions are  \begin{equation}
    \gamma_0^\prime = -r - \frac{r x_0}{h}
    \quad \text{and} \quad
    \beta_0^\prime = \frac{r x_0}{h},
\end{equation}
while the revised last row coefficients that into account the linear boundary conditions are    
\begin{equation}
    \alpha_{2^n-1}^\prime = -\frac{r x_N}{h}
    \quad \text{and} \quad
    \gamma_{2^n-1}^\prime = -r + \frac{r x_N}{h}.
\end{equation}
Consequently, for $n \geq 2$ qubits, the Black-Scholes Hamiltonian with linear boundary conditions is written as  
\begin{eqnarray}
     -i\hat{H}_{LBS}^{(n)} &=& -i\hat{H}_{BS}^{(n)} 
    + (\gamma_0^\prime-\gamma_0) \hat{A}_\nwarrow^{(n)}
    + (\beta_0^\prime - \beta_0) (\hat{A}_\nwarrow^{(n-1)} \otimes \hat{A}_\nearrow^{(1)}) \nonumber\\
    & &  +  \ (\alpha_{2^n-1}^\prime - \alpha_{2^n-1}) (\hat{A}_\searrow^{(n-1)} \otimes \hat{A}_\swarrow^{(1)})
    + (\gamma_{2^n-1}^\prime - \gamma_{2^n-1}) \hat{A}_\searrow^{(n)}.
  \end{eqnarray}

\subsection*{The rescaling protocol}
Under the assumption of linear boundary conditions,   $u(x,\tau)$ is linear with respect to $x$ in the neighbourhood of $x_0$ and $x_N$, hence, we have 
\begin{equation}
    u(x,\tau) = a(\tau)x + b(\tau)
    \label{SI-eqn:linear}
\end{equation}
for $x\approx x_0$ and $x\approx x_N$.
Substituting Eq.~(\ref{SI-eqn:linear}) into the Black-Scholes equation, Eq.~(\ref{SI-eqn:black-scholes}), we obtain  the boundary equation
\begin{align}
    \pdv{}{\tau} \left(a(\tau)x+b(\tau)\right) &= rx\pdv{}{x}\left(a(\tau)x+b(\tau)\right)  - r\left(a(\tau)x+b(\tau)\right),  
    \end{align}
which implies 
\begin{equation}
    x\dv{a}{\tau} + \dv{b}{\tau} =  -rb(\tau).\label{eqn:ODEBC}
\end{equation}
Equation~(\ref{eqn:ODEBC}) is an ODE with solutions
\begin{equation}
    a(\tau) = a(0) \quad \text{and} \quad b(\tau) = b(0)e^{-r\tau}.
\end{equation}
Determining $a(0)$ and $b(0)$ at each boundary, Eq.~(\ref{SI-eqn:linear}) becomes   a closed form expression that explains how $u(x,\tau)$ evolves at the boundaries. Assuming $h$ is sufficiently small, we can approximate $a(0)$ and $b(0)$ at each boundary by assuming that the closest sample point also follows the linear condition. Thus, on the left boundary, using $u(x_0,0) = a_0(0) x_0 + b_0(0)$ and $u(x_1,0) = a_0(0) x_1 + b_0(0)$, we have 
\begin{equation}
    a_0 = \frac{u(x_1,0) - u(x_0,0)}{h}
    \quad \text{and} \quad
    b_0(0) = u(x_0,0) - a_0 x_0. \label{eqn:lbc}
\end{equation}
Similarly, on the right boundary, we have
\begin{equation}
    a_N = \frac{u(x_N,0) - u(x_{N-1},0)}{h}
    \quad \text{and} \quad
    b_N(0) = u(x_N,0) - a_N x_N.\label{eqn:rbc}
\end{equation}
Using Eq.~(\ref{eqn:lbc}) and Eq.~(\ref{eqn:rbc}), we  define a protocol that rescales the normalised state $\ket{\bar{\psi}_u(\tau)}$ obtained from QNUTE by a factor $C_*(\tau)$ such that 
\begin{equation}
    \bra{0}C_*(\tau)\ket{\bar{\psi}_u(\tau)} = a_0 x_0 + b_0(0)e^{-r\tau},\label{eqn:lbc_c*}
\end{equation}
or
\begin{equation}
    \bra{2^n-1}C_*(\tau)\ket{\bar{\psi}_u(\tau)} = a_N x_N + b_N(0)e^{-r\tau}.\label{eqn:rbc_c*}
\end{equation}
The choice between Eq.~(\ref{eqn:lbc_c*}) and  Eq.~(\ref{eqn:rbc_c*}) depends on the preference assigned to the boundaries during simulation. This work used the  left boundary when modelling put options, and the right boundary when  modelling call options. The protocol fails if $a(0)=b(0)=0$ on both boundaries.

\end{document}